\documentclass[a4paper,journal,transmag]{IEEEtran}
\usepackage{graphicx}
\usepackage{amsmath}
\usepackage{textcomp}
\usepackage{cite}
\usepackage{siunitx}

\begin{document}

\title{A Magnetic Field Camera for Real-Time \\ Subsurface Imaging Applications}
\author{\IEEEauthorblockN{Andriyan B. Suksmono\IEEEauthorrefmark{1,2},~\IEEEmembership{Senior Member,~IEEE}, Donny Danudirdjo\IEEEauthorrefmark{1}, Antonius D. Setiawan\IEEEauthorrefmark{3}, \\ Rizki P. Prastio\IEEEauthorrefmark{1}, and Dien Rahmawati\IEEEauthorrefmark{1},~\IEEEmembership{Student Member,~IEEE}}
\medskip
\IEEEauthorblockA{\IEEEauthorrefmark{1}School of Electrical Engineering and Informatics, Institut Teknologi Bandung, Bandung 40132, Indonesia}
\IEEEauthorblockA{\IEEEauthorrefmark{2}ITB Research Center on ICT (PPTIK-ITB), Institut Teknologi Bandung, Bandung 40132, Indonesia}
\IEEEauthorblockA{\IEEEauthorrefmark{3}School of Electrical Engineering, Telkom University, Bandung 40257, Indonesia}

\thanks{
Manuscript received MMMM DD, YYYY; revised MMMM DD, YYYY; accepted MMMM DD, YYYY. Date of publication MMMM DD, YYYY; date of current version MMMM DD, YYYY. Corresponding author: A.B. Suksmono (e-mail: suksmono@stei.itb.ac.id).

Color versions of one or more of the figures in this paper are available online at http://ieeexplore.ieee.org.

Digital Object Identifier XX.XXXX/XXXX.XXXX.XXXXXXX}}

\IEEEtitleabstractindextext{
\begin{abstract}
We have constructed an imaging device that capable to show a spatio-temporal distribution of magnetic flux density in real-time. The device employs a set of AMR (Anisotropic Magneto Resistance) 3-axis magnetometers, which are arranged into a two-dimensional sensor array. All of the magnetic field values measured by the array are collected by a microcontroller, which pre-process and send the data to a PDU (Processing and Display Unit) implemented on a smartphone/tablets or a computer. An interpolation algorithm and display software in the PDU present the field as a high-resolution video; hence, the device works as a magnetic field camera. In the experiments, we employ the camera to map the field distribution of distorted ambient magnetic field induced by a hidden object. The obtained image of field shows both the position and shape of the object. We also demonstrate the capability of the device to image a loaded powerline cable carrying a 50~Hz alternating current.
\end{abstract}
\begin{IEEEkeywords}
magnetic camera, magnetic imaging, subsurface imaging, digital compass, magnetometer, bilinear, bicubic, AMR (anisotropic magneto-resistance), GMR (giant magneto-resistance)
\end{IEEEkeywords}}

\maketitle

\section{Introduction}

\IEEEPARstart{C}AMERA is an image-capturing device. Early camera uses chemical substance to record image of an object. In principle, light rays reflected by the object are collected by a lens or a pinhole to form a small planar image, which falls on a negative film that contains a layer of photo-sensitive chemicals. In a photographic development process, the negative film is converted into a positive one on a paper. 
Modern cameras record the image digitally. The object image that has been projected onto a recording area is measured by a light-sensitive sensor array. The voltage values from the array are then digitized so that an array of numbers representing the image of the object can be stored in a digital format or displayed on a screen. Specialized cameras are usually named after the type of waves they capture, such as infra-red, X-ray, or hyper-spectral cameras.

A distinctive feature of a camera is that all picture elements of the object image (or sequence of images/video) are captured simultaneously, instead of elements-by-elements or pixel-by-pixel. In the latter case, the device is usually called a scanner. A digital camera can also display the captured image instantly. Considering these features, in this paper we will refer to the generic name \emph{camera} as \emph{a device that capable to capture and display a distribution of physical quantity of an object instantly}. For optical cameras, the physical quantity is the reflectivity or physical response of the object from impinging lights (including the colors). 

In \cite{suksmono2017}, we have constructed a magnetic field imaging system utilizing the built-in magnetometer of a smartphone. To obtain an image that represents distribution of magnetic flux density induced by ambient magnetic field; such as the earth magnetism \cite{larmor1919}, one has to scan an imaging area and then execute a reconstruction algorithm to obtain entire field values on the area. Therefore, this device is categorized as a scanner, which will be referred to as \emph{B-Scanner} where \emph{B} in the name follows the notation of magnetic flux density, which is normally denoted by $\vec{B}$. Magnetic field scanners have also been employed in geomagnetic survey \cite{mariita2007} and industrial testing \cite{orozco2013}. Micro-circuitry faults of an IC (Integrated Circuits) can also be detected by magnetic field scanner \cite{knauss2004}. A close relative to the magnetic imaging is the eddy-current imaging, which also has been investigated actively by researchers \cite{mccary1984, tsukada2006, volk2008, joubert2013, bore2015}.

In magnetic field surveys, one typically uses a highly sensitive sensor solely designed for a research purpose \cite{mariita2007}. Nowadays, magnetic field sensors are easily found at low prices. These sensors make use of giant magneto resistance (GMR) \cite{fert1988, grunberg1989, moodera1995} or anisotropic magneto resistance (AMR) phenomena, which is the change in resistance of a material due to exerting magnetic field. In contrast to the conventional sensor, such as the proton magnetometer, the GMR/AMR based devices can be implemented as a compact and low cost IC sensors. Most mobile phones today include a built-in magnetometer as one of their standard features. 

This paper describes a design and realization of a magnetic field camera or the \emph{B-Camera}. The proposed B-Camera has a capability to capture and display magnetic field distribution of a region instantly, thanks to an array of magnetometers that measures the field values on a regular grid of the area simultaneously. Then, a reconstruction algorithm interpolates entire values within the observed area and display the result on a screen. We use a similar reconstruction and display subsystems as in the B-Scanner; i.e, data collected by the sensor-array are send to a PDU (processing and display unit); which can be implemented on smartphones, tablets, or computers, through a USB or a Bluetooth port.

The rests of the paper are organized as follows. In Section II, we briefly describe the principle of the proposed B-Camera. The following Section III describes the experiments, including the arrangement, reconstruction of image, and analysis of the results. Section IV summarizes and concludes the paper.

\medskip{}
\section{Materials and Methods}

\subsection{B-Camera: A Magnetic Field Camera}
Human eyes cannot sense the presence of magnetic field. To "see" the magnetic field, one can use iron powder. By pouring the powder onto a sheet of paper which is placed on a top of a magnet, we obtain a picture of magnetic field lines. The flux density of the field is indicated by the density of the lines. More accurate measurement requires the use of a magnetometer, which is usually brought around to scan an area while the field values are recorded. The field distribution can be derived by interpolating to locations within the surveyed area where the values are unknown. For an instantaneous display of the field---as required in a camera, an array of magnetometers is necessary. In \cite{tuan2014}, a magnetic camera made of Hall-effect sensor has been proposed where magnetic field strength of observed object are captured and displayed as a three-dimensional surface map.

Additionally, we also cannot do focusing to a magnetic field, which normally done in an optical camera by using a lens. Therefore, the field is sensed directly at the corresponding position by a magnetometer. The {B-scanner} in \cite{suksmono2017} operates using a predefined grids over an observed area and measures the field at the center of each grid. Dividing the area into grids are equivalent to discretization of the space, whereas the magnetometer of a smartphone in the {B-Scanner} also gives discretized values of the magnetic field. Simultaneous observation of the field values on the area can be obtained by placing a number of sensors at the center of the grids, which yields an array of magnetometers. At this point, there is a close similarity between the B-Camera and an optical camera.

\begin{figure}
	\includegraphics[width=1\columnwidth]{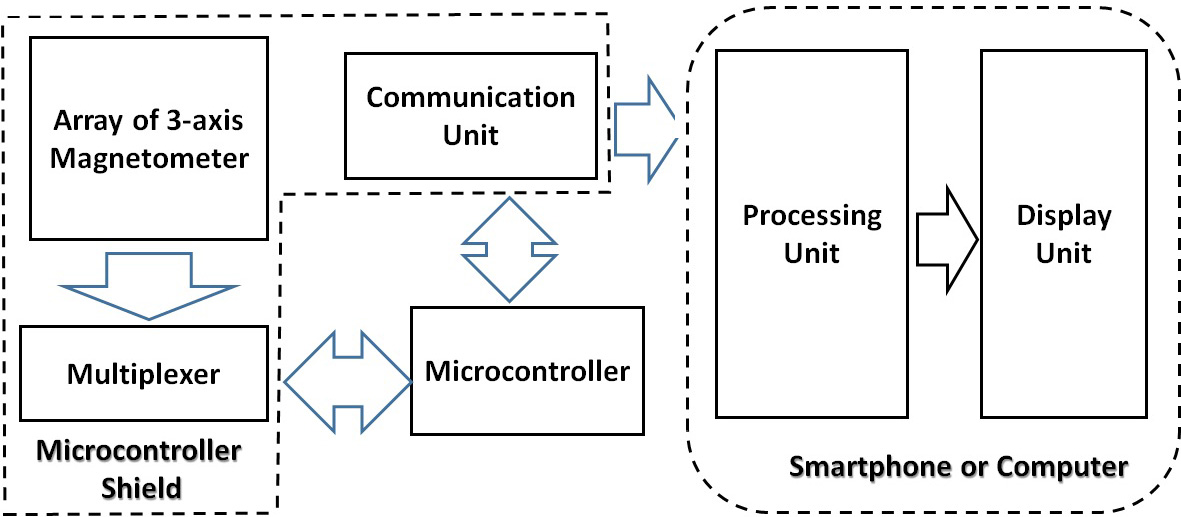}
	\protect\caption{\label{LabelFig1}A block diagram of the B-Camera showing its subsystems and their interactions. It consist of an Array of magnetometers and Multiplexer (implemented as an Arduino shield), a Microcontroller Unit (Arduino), a Communication Unit (Bluetooth or USB port), and a Processing and Display Unit (implemented in a smartphone or a computer).
	}
\end{figure}

\subsection{Construction of The B-Camera}
In principle, the B-Camera works like the magnetic field scanner, unless the sensors position are fixed at a regular grid points/array. A block diagram displayed in Fig.\,1 shows the construction of the {B-Camera}. 

\subsubsection{Sensor Array and Multiplexer}
The sensor-array performs spatial sampling of the magnetic field on an area, which is achieved by arranging the magnetometers into a two-dimensional array. The minimum spatial sampling size, or the size of a sampling grid, will be related to the dimension of the sensor. Since the present implementation uses a breakout of an HMC5883L magnetometer chip, we set the grid cell size to be \SI[mode=text]{2}{{cm}^2}. The sensors are arranged into \num[mode=text]{4x4} array, so that the total size of the sensed area is \num[mode=text]{8x8}~\SI[mode=text]{}{{cm}^2}. 

The HMC5883L sensor can measure magnetic field up to 8~gauss in three axis. The chip employs I2C (IC-to-IC Communication) protocol to communicate with other device, which in this case is a microcontroller. Since all of the sensors have identical fixed address, we employed a 1-to-16 demultiplexer to access the field values of each sensor, which is controlled by four digital channels of the microcontroller. This way, the measured field values can be read sequentially by the microcontroller from each sensor's data registers.

\subsubsection{Microcontroller Unit}
The main tasks of the Microcontroller Unit are to read the sensors, to do a basic processing, and to send the result to the PDU. We have employed an Arduino microcontroller because of its availability, affordability, and its openness. We read the data through four digital channels of the microcontroller and translate the integer (scaled) field values into a re-scaled proper values in gauss unit. The results are then sent to PDU by using a USB port, if the PDU is implemented on a computer, or by using a Bluetooth channel for the smartphone-based PDU.

\subsubsection{Communication Unit}
We can use either USB port or Bluetooth for communication between the sensor array and the processing unit. When mobility is prioritized, we can use a Bluetooth breakout such as the HC-06. The array data includes the ID of each sensor, the three components of measured magnetic fields  $\{B_x, B_y, B_z \}$, and a flag to indicate the end of a scan, so that the PDU can immediately process the data and display the image when all of the data in the array have already been received.

\subsubsection{PDU (Processing and Display Unit)}
The PDU can be implemented on either a computer or a smartphone/tablet. Present days smartphones, despite of its affordability, have a sufficient computing power and display resolution to be used for this purpose. The main task of the computing part is to interpolate the two-dimensional magnetic field values into a larger size. The purpose of interpolation is to present the user with a high-resolution  image/video. We have implemented two interpolation algorithms in our device, i.e. bilinear and bicubic \cite{suksmono2014, suksmono2017}. It has been coded in Java using Android Studio for the smartphone based PDU and by using Matlab for the computer based PDU.

The bilinear method requires four known values obtained by measurements to estimate a field value at a particular point. Based on these known values, the values on desired positions are derived by using linear interpolation. The drawback of the bilinear method is, when the sampling points are sparsely distributed, the constructed field cannot be smoothly interpolated. To obtain a better result, we can use bicubic method that uses 16 neighboring values to estimate unknown values. The interpolation is done by nonlinear function, therefore, a smoother result can be obtained. 

The display unit shows magnetic field distribution for each components and the total magnitude. For a given measured field at a discrete point $(m,n)$ within the surveyed area, where $m$ and $n$ indicates the row and column number, respectively, the vector of magnetic field is given by
\begin{equation}
  \label{EQ_Bvec}
  \vec{B}(m,n) = B_x(m,n)\hat{i} + B_y(m,n)\hat{j} +  B_z(m,n)\hat{k}    
\end{equation}
where $\hat{i},\ \hat{j},\ \hat{k} $ are unit vectors to the direction of $x$, $y$, and $z$ respectively. The magnitude of the magnetic flux density is given by
\begin{equation}
  \label{EQ_Bmag}
  |\vec{B}(m,n)| = \sqrt{B_x^2(m,n) + B_y^2(m,n) +B_z^2(m,n)}    
\end{equation}
Each of the components $\{B_x,\ B_y,\ B_z\}$ and the magnitude $|B|$ are displayed separately by the PDU system. Observation of field components is hardly possible in other imaging modalities, such as radar and optical imaging. It enriches the features which can increase  recognition capability of the system.

\begin{figure}
	\begin{center}
	  \includegraphics[width=1\columnwidth]{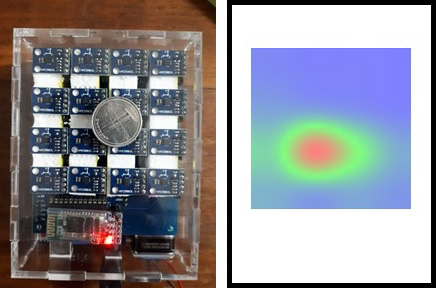} \\[1mm]
	  \footnotesize{(a)}\\[3mm]
	  \includegraphics[width=1\columnwidth]{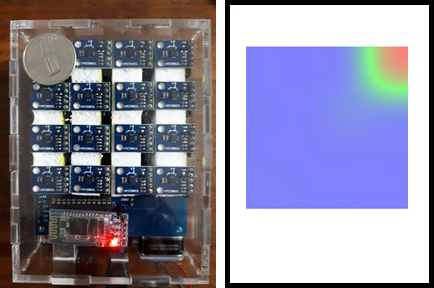} \\[1mm]
	  \footnotesize{(b)}
	\end{center}
	\protect\caption{\label{LabelFig3}Position of an object at B-Camera's array and their corresponding magneto-photograph. The array is designed to be used downward, so that the left-right position of the image has been inverted. The object is a Rp.1000 nickel coin: (a) it is positioned at the center of the the array, (b) it is located at the corner of the array. Images on at the right part of (a) and (b) shows the consistency of the field distribution with their corresponding locations at the left part.}
\end{figure}

\medskip{}
\section{Experiments and Analysis}
\subsection{Imaging A Coin with a Smartphone-Based PDU}
In this experiment, we have used the B-Camera with smartphone based PDU. The object was a nickel coin (Rp.1000), which possesses a ferromagnetic property. In the first experiment, the coin was located at the center of the array, as shown in the left part of Fig.~\ref{LabelFig3}(a). The magnitude field distribution expressed in Eq.\eqref{EQ_Bmag} was displayed on a smartphone screen shown in the right part of Fig.\ref{LabelFig3}(a). The image shows a circular distribution of the field, which is strong at the center of the array. This image is consistent with the position and shape of the object.

In the second experiment, the coin was located at the corner of the array. It should be noted that in a normal usage, the array looks downward; therefore, the left-right positions are exchanged. Left part of Fig.~\ref{LabelFig3}(b) shows the position of the object on the array, whereas the right part shows the field distribution. This figure shows image of the object as expected, indicating that the device have worked properly. 

\subsection{Observation of a Hidden Object}
In this experiment, an object (a 9~V dry-cell battery) was located on a top of a book stack. We have used B-Camera with a computer based PDU. The microcontroller (Arduino) was connected to a USB port of the computer, where a Matlab program was running. By using the Matlab, we can obtain a better color contrast easily, which enhanced the display of the magnetic field distribution. The purpose of this experiment was to show that a hidden object separated by a thick wall can be detected (and imaged) by the B-Camera. 

\begin{figure*}
	\begin{center}
		\includegraphics[width=1\textwidth]{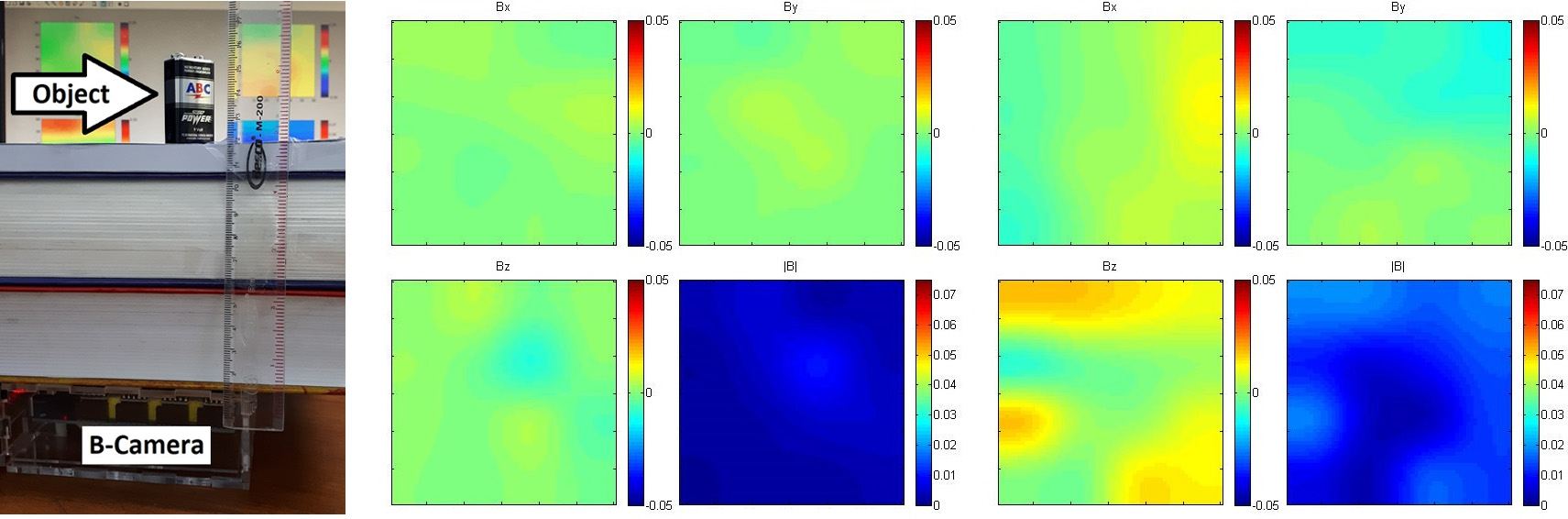}
	\\[2mm]
	\footnotesize{(a)}\hspace{53mm}
	\footnotesize{(b)}\hspace{67mm}
	\footnotesize{(c)}\hspace{13mm}\,
	\end{center}\vspace{-2mm}
	\protect\caption{\label{Label_DoP} Experiments to show the capability of the B-Camera to image a hidden object: (a) A dry-cell battery, whose cover is made of iron, is located on top of a book stack to simulate a \SI[mode=text]{12}{cm} wall hiding an object, (b) Magnetic field distribution with subtracted background without an object, (c) Field distribution after the object has been put on top of the book stack. Color bar in (b) indicates magnetic field strength in gauss unit. By comparing figure (b) to (c), we observe that the field distribution has changed. It shows that the device is capable to detect (and image) ferromagnetic objects hidden behind a thick wall.}
\end{figure*}

Fig.\,\ref{Label_DoP} shows (a) experiment setting (test-range), (b) field distribution before an object was put, (c) field distribution after the object has been placed on top of a book stack. The ruler in Fig.\,\ref{Label_DoP}\,(a) indicates that the distance between the sensor array to the object is around \SI[mode=text]{12}{cm}. By comparing the field distribution without the object in Fig.\,\ref{Label_DoP}\,(a) with the one with an object shown in Fig.\,\ref{Label_DoP}\,(b), we can see that the field has been changed and therefore the B-camera was capable to sense the field of an object located behind a thick wall. The colors represent the field strength in gauss (G), whose values are indicated in the color bar.

\subsection{Imaging A Loaded Powerline Cable}
In this experiment, we have measured the magnetic field induced by current in a loaded 50~Hz powerline cable. Since the cable (wire) was much longer than the length of the array, the induced magnitude of magnetic field strength $|\vec{B}(m,n)| \equiv B(m,n)$ by a current $I$ at an array sensor $(m,n)$ located at a distance $r(m,n)$ from the center of the cable  is given by
\begin{equation}
  \label{EQ_Bwire}
    B(m,n) =\frac{\mu_0 I}{2 \pi r(m,n)}  
\end{equation}
where $\mu_0$ is the permeability of the air (free space). The field strength is proportional to the current $I$ and inversely proportional to the distance between the wire and the sensor. Since the array is planar, whereas the cable is located above the center of the array, the field will be stronger at the center of the array and weaker at the edge. 

\begin{figure}
    \begin{center}
	\includegraphics[width=1\columnwidth]{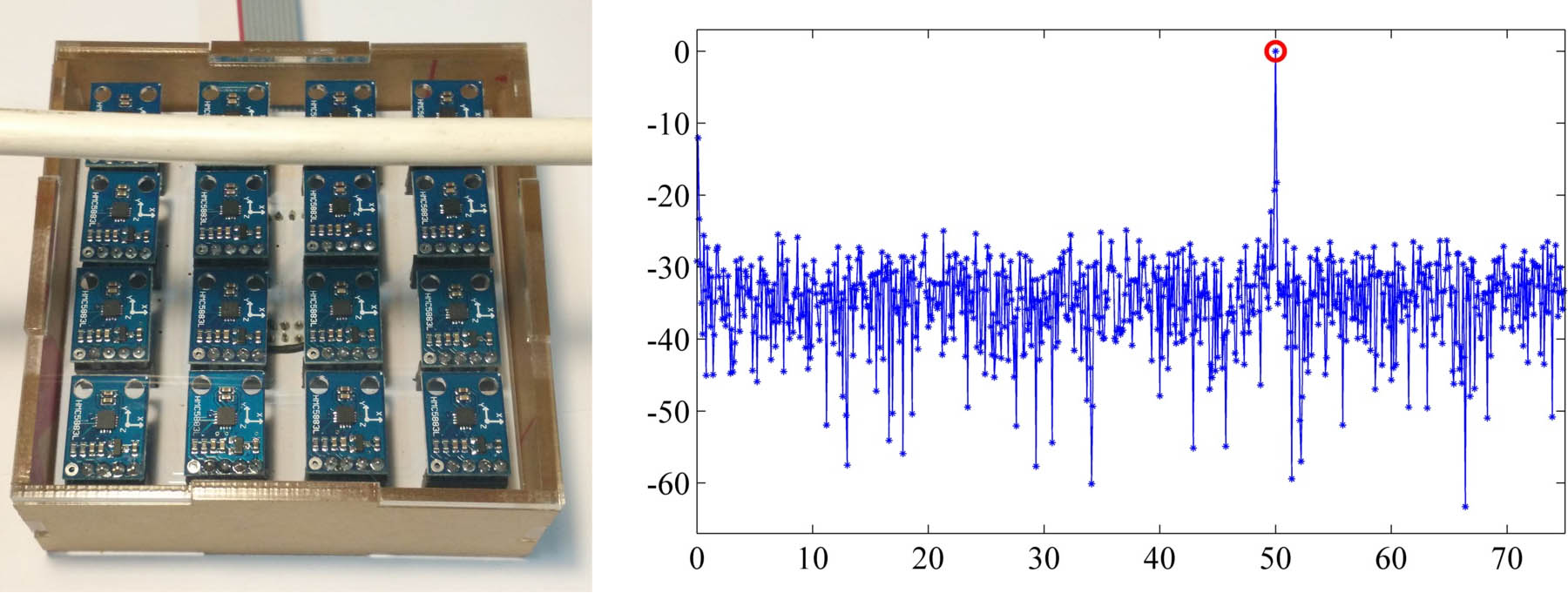} \\[2mm]
	\footnotesize{(a)} \\[4mm]
	\includegraphics[width=1\columnwidth]{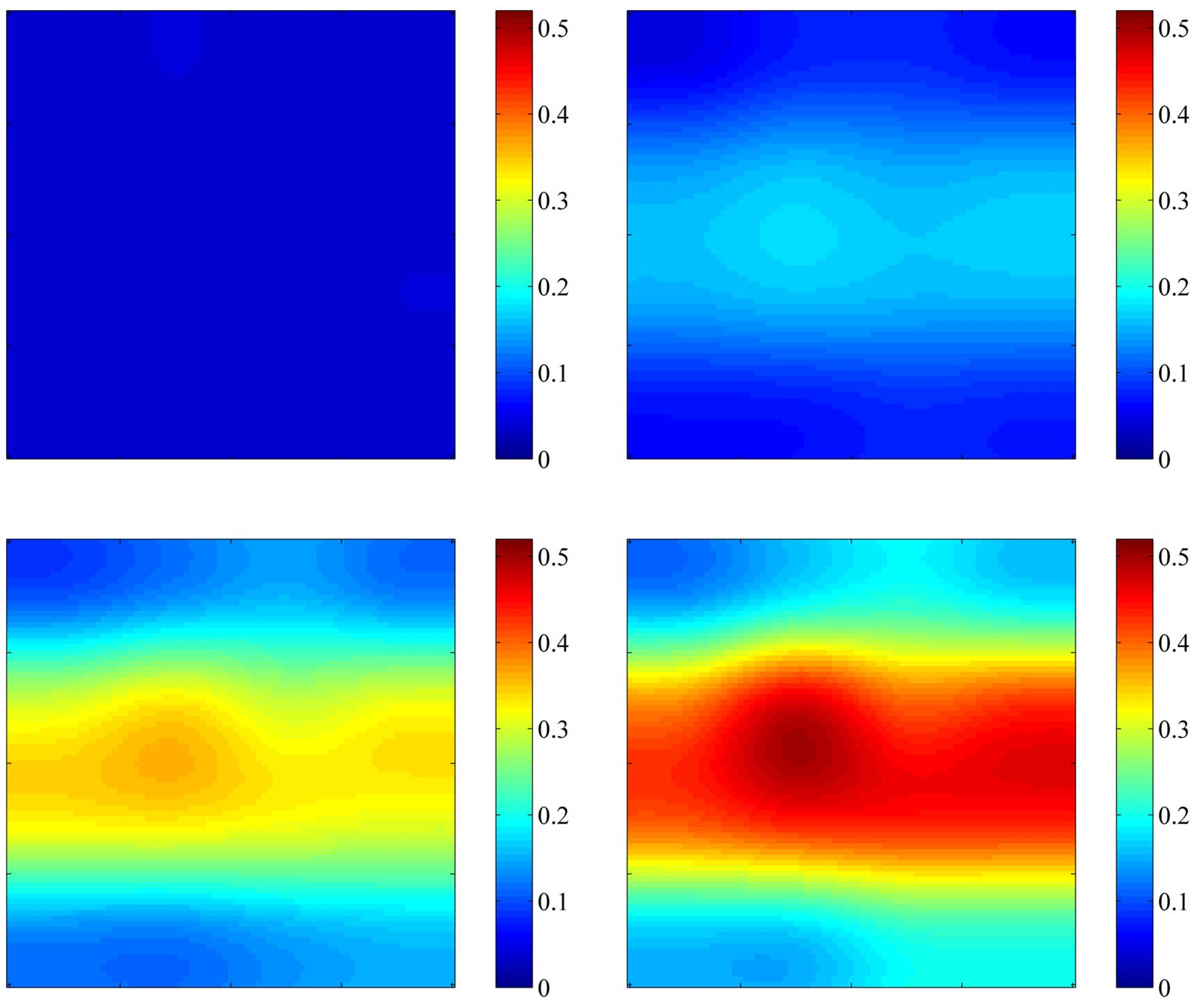} \\[2mm]
	\footnotesize{(b)} \\[1mm]
	\protect\caption{\label{FigPowerline} Imaging a loaded 50~Hz powerline cable. (a) Left part shows the experiment setup where the cable is located around 3 cm from the B-Camera. Right part shows the spectrum of the captured data. (b) Magnitude of magnetic flux density of the 50~Hz components: (upper left) when the cable is not loaded, (upper right) with a 25~W load, (lower left) with a 40~W load, and (lower right) with a 60~W load.}
	\end{center}
\end{figure}

Fig.~\ref{FigPowerline} shows (a) the experiment setup and spectrum of the data, and (b) the observed results for various loads. The sampling rate of the magnetometer is set to 150 Hz for all field components. Spectrum of the original data displayed in the right part of Fig.~\ref{FigPowerline}(a) shows a peak at 50~Hz. The collected data were filtered, so that only the 50~Hz component of the data were extracted.

Magnitude distribution of the magnetic fields for four loads, i.e. \SI[mode=text]{0}{W}, \SI[mode=text]{25}{W}, \SI[mode=text]{40}{W}, and \SI[mode=text]{60}{W} incandescent lamps, are displayed in Fig.~\ref{FigPowerline}(b). The figures shows that the flux density increases with the increasing power dissipation of the loads, which can be understood since for a given voltage $V$, the current $I$ is proportional to the power. We also observed that the flux density is strong at the center and weaker at the edge of the array. These results are consistent with Eq.(\ref{EQ_Bwire}).

\section{Conclusions and Further Directions}
A magnetic field camera has been constructed. The array sensors are built from digital magnetometers, consisting of \num[mode=text]{4x4} elements. The camera has been demonstrated to map the magnetic field distribution of ferromagnetic objects behind a surface, and to display the distribution at an instant time. Additionally, the high sampling rate of the sensors also allows the camera to measure and map the distribution of magnetic field near a loaded 50~Hz powerline cable.

\section*{Acknowledgements}
This work has been supported by the ITB-Asahi Glass Foundation Grant of Research 2018.

\bibliography{Referensi_BCamera}
\bibliographystyle{ieeetr}

\end{document}